\documentstyle[12pt,epsfig,axodraw,a4]{article} 
\def\bild#1#2{    
        \vspace*{-5mm}
        \begin{center}
        \begin{math}
        \epsfxsize#2cm
        \epsffile{#1}
        \end{math}
        \end{center}  }
\newcommand{\vs}{\vspace{-0.25cm}}
\begin{document} 
\begin{center}
\large{\bf Spectral functions of isoscalar scalar and isovector 
electromagnetic form factors of the nucleon at two-loop order}

\bigskip

N. Kaiser\\

\medskip

{\small Physik Department T39, Technische Universit\"{a}t M\"{u}nchen,
    D-85747 Garching, Germany}

\end{center}

\medskip

\begin{abstract}
We calculate the imaginary parts of the isoscalar scalar and isovector 
electromagnetic form factors of the nucleon up to two-loop order in chiral 
perturbation theory. Particular attention is paid on the correct behavior of 
Im\,$\sigma_N(t)$ and Im\,$G_{E,M}^V(t)$ at the two-pion threshold $t_0=4
m_\pi^2$ in connection with the non-relativistic $1/M$-expansion. We recover 
the well-known strong enhancement near threshold originating from the nearby 
anomalous singularity at $t_c = 4m_\pi^2-m_\pi^4/M^2 = 3.98 m_\pi^2$. In 
the case of the scalar spectral function Im\,$\sigma_N(t)$ one finds a
significant improvement in comparison to the lowest order one-loop result. 
Higher order $\pi\pi$-rescattering effects are however still necessary to close
a remaining 20\%-gap to the empirical scalar spectral function. The isovector
electric and magnetic spectral functions Im\,$G_{E,M}^V(t)$ get additionally
enhanced near threshold by the two-pion-loop contributions. After supplementing
their two-loop results by a phenomenological $\rho$-meson exchange term one can
reproduce the empirical isovector electric and magnetic spectral functions
fairly well. 
\end{abstract}


\bigskip

%To be published in: {\it The Physical Review C (2003), Brief Report}

\bigskip 

\bigskip

The structure of the nucleon as probed in elastic electron-nucleon scattering
is encoded in four form factors $G_{E,M}^{S,V}(t)$ (with $t$ the squared 
momentum transfer). The understanding of these form factors is of importance 
for any theory or model of the strong interactions. Abundant data on the 
electromagnetic form factors of the nucleon exist over a large range of 
momentum transfers, $0< \sqrt{-t} < 6\,$GeV. Dispersion theory is a tool to 
interpret these data in a largely model-independent fashion. Each 
electromagnetic form factor $G_{E,M}^{S,V}(t)$ can be written in terms of an 
unsubtracted dispersion relation and their imaginary parts are often modeled 
by a few vector meson poles. This procedure is based on the successful vector 
meson dominance hypothesis which states that the (virtual) photon couples to 
hadrons only via intermediate vector mesons. However, as pointed out already in
1959 by Fulco and Frazer \cite{fulco} such an approach is not in conformity 
with general constraints from unitarity and analyticity. In particular, the 
singularity structure of the triangle diagram (left graph in Fig.\,1) is not 
respected at all when using only vector meson poles. As a consequence of a 
nearby logarithmic singularity at $t_c = 4m_\pi^2-m_\pi^4/M^2$ (on the second
Riemann-sheet) the low-mass two-pion continuum has a very pronounced effect on 
the isovector electric and magnetic spectral functions Im\,$G_{E,M}^V(t)$ on 
the left wing of the $\rho(770)$-resonance. This becomes particularly visible 
in the isovector mean square radii of the nucleon. The effect was quantified by
H\"ohler and Pietarinen \cite{hoehler} in their work on the nucleon 
electromagnetic form factors based on $\pi\pi\to \bar NN$ partial wave 
amplitudes. The dispersion relation analysis of the nucleon electromagnetic 
form factors was recently refined by the Bonn-Mainz group \cite{mergel,hammer}
accounting for new electron scattering data and the high-energy constraints
from perturbative QCD. 

The first one-loop calculation of the isovector electromagnetic form factors
has been performed in ref.\cite{gss} using the fully relativistic version of 
baryon chiral perturbation theory. The correct analytical structure naturally
emerged in this calculation and thus the strong enhancement near the two-pion
threshold. In the framework of heavy baryon chiral perturbation theory 
next-to-leading order one-loop results for the isovector electric and magnetic 
spectral functions Im\,$G_{E,M}^V(t)$ have been given in ref.\cite{spectral}. 
To that order a formally incorrect threshold behavior, not following the 
required p-wave phase space Im\,$G_{E,M}^V(t) \sim (t-4m_\pi^2)^{3/2}$, has 
been obtained. This deviation is a consequence of the coalescence of the normal
threshold $t_0=4m_\pi^2$ and the anomalous threshold $t_c = 4m_\pi^2-m_\pi^4/
M^2$ in the heavy nucleon limit $M\to \infty$. In that paper \cite{spectral} 
also the isoscalar electric and magnetic spectral functions Im\,$G_{E,M}^S(t)$
(starting at two-loop order) have been computed. It has been shown that no 
substantial contribution arises from the three-pion continuum below the 
narrow $\omega(782)$-resonance.

The purpose of this paper is to present results for the isovector 
electromagnetic spectral functions up to two-loop order in chiral perturbation
theory paying special attention on the correct threshold behavior. We will 
include in our discussion also the isoscalar scalar form factor of the nucleon.
Although not directly measurable this form factor $\sigma_N(t)$ is very 
interesting since it quantifies explicit chiral symmetry breaking effects in 
the nucleon ($\sigma_N(0)= \hat m\,\partial M /\partial \hat m $ is the
celebrated pion-nucleon sigma-term). Moreover, its spectral function 
Im\,$\sigma_N(t)$ is completely dominated by the low-mass two-pion continuum
due to the abovementioned threshold enhancement effect with no visible scalar
resonance structures (see Fig.\,4 in ref.\cite{sigmaterm}).  
  
\bigskip

\bild{spectr1.epsi}{13}
{\it Fig.1: One-loop diagrams contributing to the imaginary parts of the 
isoscalar scalar and isovector electromagnetic form factors of the nucleon.
Dashed and solid lines denote pions and nucleons, respectively. The wiggly 
line represents the external scalar or vector source. The heavy dot symbolizes 
the second-order chiral $\pi\pi NN$-contact vertex proportional to the 
low-energy constants $c_{1,2,3,4}$. The combinatoric factor of the last two 
diagrams is 1/2.}

\bigskip

Let us begin with recalling the basic definition of the isoscalar scalar form 
factor $\sigma_N(t)$ of the nucleon: 
\begin{equation} \langle N(p')|\hat m \bar q q|N(p)\rangle 
= \sigma_N(t) \, \bar u(p') u(p) \,. \end{equation}
Here, $\hat m=(m_u+m_d)/2$ denotes the average light quark mass, $t=(p'-p)^2$ 
is the Lorentz-invariant squared momentum transfer and $u(p)$ stands for a
Dirac spinor. Similarly, the matrix element of the isovector vector current
defines the isovector Dirac and Pauli form factors:
\begin{equation} \langle N(p')|\bar q \gamma^\mu{\tau_a \over 2}q|N(p)\rangle 
= \bar u(p') \bigg[ \gamma^\mu F_1^V(t) +{i\sigma^{\mu\nu} \over 2M}(p'-p)_\nu 
\,F_2^V(t)\bigg] \tau_a u(p) \,, \end{equation}
where $M=938.27\,$MeV denotes the proton mass. In what follows, we will work
with other linear combinations of $F_{1,2}^V(t)$, the isovector electric and
magnetic (Sachs) form factors: 
\begin{equation} G_E^V(t) = F_1^V(t) + {t \over 4M^2} F_2^V(t)\,, \qquad
G_E^V(0) = {1\over 2} \,, \end{equation}
\begin{equation} G_M^V(t) = F_1^V(t) + F_2^V(t)\,, \qquad
G_M^V(0) = {\mu_p-\mu_n\over 2} = 2.353 \,, \end{equation}
which obey at $t=0$ the normalization conditions written in eqs.(3,4). As
usual, $\mu_p= 2.793$ and $\mu_n = -1.913$ denote the proton and neutron
magnetic moments (in units of nuclear magnetons). In the complex $t$-plane the
form factors  $\sigma_N(t)$ and $G_{E,M}^V(t)$ have a right hand cut along the
positive real axis starting at $t_0=4m_\pi^2$ related to the opening of the 
two-pion threshold. The imaginary parts Im\,$\sigma_N(t+i0^+)$ and
Im\,$G_{E,M}^V(t+i0^+)$ of these nucleon form factors (equal to the half
discontinuities across the cut) are also called their spectral functions.

The contributions to the spectral functions Im\,$\sigma_N(t)$ and Im\,$G_{E,M
}^V(t)$ calculated up to two-loop order in chiral perturbation theory can be 
grouped into four different classes. The first class of contributions comes
from the one-loop diagrams with leading order relativistic $\pi N$-interaction
vertices and propagators (represented by the first and second diagram in
Fig.\,1). For these one-loop graphs we can take the fully relativistic
expressions given in ref.\cite{gss} and expand them in inverse powers of the
large nucleon mass $M$ up to order ${\cal O}(M^{-2})$. This way one gets for
the imaginary part of the isoscalar scalar form factor of the nucleon:    
\begin{equation} {\rm Im}\,\sigma_N(t)={3g_A^2 m_\pi^2 \over 64\pi f_\pi^2
\sqrt{t}} \bigg\{(t-2m_\pi^2) A(t)\bigg[1+{3t \over 8 M^2}\bigg] -{Q \,t\over 
2M} \bigg\}\,,  \end{equation}
with the auxiliary functions: 
\begin{equation}  A(t) = \arctan{Q \sqrt{4M^2-t}\over t-2m_\pi^2} 
\,, \qquad Q= \sqrt{t-4m_\pi^2}\,. \end{equation}
The same procedure gives for the imaginary parts of the isovector electric and
magnetic form factors of the nucleon:
\begin{eqnarray} {\rm Im}\,G_E^V(t)&=&{1\over 64\pi f_\pi^2\sqrt{t}} \bigg\{ 
{Q \over 3}\Big[g_A^2(5t-8m_\pi^2)+Q^2\Big] \nonumber \\ && -{g_A^2 \over M} 
(t-2m_\pi^2 )^2 A(t) \bigg[1+{3t \over 8 M^2}\bigg] + {g_A^2 Q \,t \over 2M^2}
(t-2m_\pi^2 ) \bigg\}\,,  \end{eqnarray}
\begin{eqnarray} {\rm Im}\,G_M^V(t)&=&{M\over 32\pi f_\pi^2\sqrt{t}} \bigg\{ 
g_A^2 Q^2 A(t) +{Q \over 6M} \Big[ g_A^2 (10m_\pi^2-4t)+Q^2 \Big] \nonumber \\
&&+{g_A^2\over 8M^2} (3t^2-12t m_\pi^2 +8m_\pi^4) A(t) \bigg\}\,.\end{eqnarray}
It is important to note that the expressions in eqs.(5,7,8) possess the correct
threshold behavior, namely Im\,$\sigma_N(t)\sim \sqrt{t-4m_\pi^2}$ and Im\,$
G_{E,M}^V(t) \sim (t-4m_\pi^2)^{3/2}$. In order to achieve this property in all
three cases a subleading term proportional to $M^{-3}$ had to be kept in
the isovector electric spectral function Im\,$G_E^V(t)$. One easily verifies
that the coefficients of $Q=\sqrt{t-4m_\pi^2}$ appearing in the threshold 
expansion of Im\,$G_{E,M}^V(t)$ vanish to the respective order in the $1/M
$-expansion. The crucial point about the representations eqs.(5,7,8) is not to 
further expand the function $A(t)$ in powers of $1/M$ as it is done 
(implicitly) in heavy baryon chiral perturbation theory \cite{spectral}. In 
fact this function $A(t)$ incorporates the anomalous (logarithmic) singularity 
at $t_c = 4m_\pi^2-m_\pi^4/M^2$, determined as that $t$-value where the 
argument of the arctangent-function becomes equal to $\pm i$. As a matter of 
fact the $1/M$-expansion of the function $A(t)$ would destroy the correct 
analytical structure and moreover generate singular terms of the form 
$Q^{-n}$. From the point of view of chiral power counting the function $A(t)$
resumes an infinite string of terms starting at zeroth order. We note aside
that the one-loop calculation of ref.\cite{infrared} employing a so-called
infrared regularization scheme gives naturally the correct threshold behavior
of Im\,$G_{E,M}^V(t)$ since there the spectral functions are identical to those
of the fully relativistic framework \cite{gss}.    

The next class of contributions is given by the one-loop diagrams with one
second order chiral $\pi\pi NN$-contact vertex proportional to the low-energy
constants $c_{1,2,3,4}$ \cite{nadja}. We obtain from the last diagram in
Fig.\,1 the following contributions to the spectral functions:
\begin{equation} {\rm Im}\,\sigma_N(t)={m_\pi^2 \,Q\over 64\pi f_\pi^2\sqrt{t}}
\Big[4m_\pi^2(c_2+3c_3-6c_1)-t(c_2+6c_3)\Big]\,,  \end{equation}
\begin{equation}{\rm Im}\,G_E^V(t) = {c_4 \,Q^3 \sqrt{t} \over 192 \pi M
f_\pi^2} \,, \qquad {\rm Im}\,G_M^V(t) = {c_4 M \,Q^3 \over 48\pi f_\pi^2
\sqrt{t}}\,, \end{equation}
where we have included also the first relativistic $1/M$-correction (as far as
it does not vanish). The separation of the $\pi\pi NN$-contact vertex into an 
isoscalar part proportional to $c_{1,2,3}$ and an isovector part proportional 
to $c_4$ is obvious from eqs.(9,10). The same $c_4$-contribution to
Im\,$G_M^V(t)$ has been written down first in eq.(10) of ref.\cite{spectral}. 

\bigskip

\bild{spectr2.epsi}{15}
{\it Fig.2: Two-loop diagrams contributing to the imaginary parts of the 
isoscalar scalar and isovector electromagnetic form factors of the nucleon.
The grey disk symbolizes all one-loop diagrams of elastic $\pi N$-scattering. 
The combinatoric factor of the first two diagrams is 1/2. The vertical dotted 
line indicates that cut which splits off the (one-loop) scalar or charge form 
factor of the pion from these factorizable diagrams.}

\bigskip

Now we turn to the two-loop diagrams. These are symbolically represented by the
left graph in Fig.\,2. The grey disk should be interpreted such that it 
includes all one-loop diagrams of elastic $\pi N$-scattering. Exploiting
unitarity the imaginary parts Im\,$\sigma_N(t)$ and Im\,$G_{E,M}^V(t)$ can 
calculated from these two-loop diagrams as integrals of the product of 
$scalar/vector\,\, source \to 2\pi$ and $2\pi \to \bar NN$ transition amplitude
over the Lorentz-invariant two-pion phase space. When evaluated in the $\pi\pi$
center-of-mass frame the pertinent two-pion phase space integrals become 
proportional to simple angular integrals $\int_{-1}^1 dx$. Consider now that 
cut which splits a two-loop diagram into a tree-level $source \to 2\pi$ 
coupling and a one-loop diagram of elastic $\pi N$-scattering. Putting together
all such terms we find the following two-loop contribution to the imaginary 
part of the scalar form factor of the nucleon:
\begin{eqnarray} {\rm Im}\, \sigma_N(t) &=& {3m_\pi^2 \over \pi^2 (8f_\pi)^4} 
\bigg\{ {4m_\pi^2\over \sqrt{t}} \Big[g_A^4(10m_\pi^2-4t)-m_\pi^2 \Big] \ln{
\sqrt{t} +Q\over 2 m_\pi} \nonumber \\ && +Q \bigg[\bigg({1\over 2}-g_A^4\bigg)
(t-2m_\pi^2) +{ 8g_A^4 m_\pi^2(\sqrt{t}-2m_\pi)(5\sqrt{t}+8m_\pi)\over 3(t+
2m_\pi \sqrt{t})} \nonumber \\ && + g_A^2 (m_\pi^2-2t) \bigg({4m_\pi \over
\sqrt{t}} + {2m_\pi^2-t \over t} \ln{\sqrt{t} +2m_\pi \over \sqrt{t}-2m_\pi} 
\bigg) \bigg] \bigg\}\,.\end{eqnarray}
This expression is identical to the $2\pi$-phase space integral $3m_\pi^2(32\pi
\sqrt{t})^{-1} \int_{-Q/2}^{Q/2} dk\,{\rm Re}\,g^+(ik,t)$ with $g^+(\omega,t)$ 
the isospin-even non-spin-flip one-loop $\pi N$-amplitude \cite{nadja,pin}. The
analogous contribution to the isovector electric spectral function reads:
\begin{eqnarray} {\rm Im}\, G_E^V(t) &=& {g_A^4m_\pi^2(2m_\pi^2-t)\over 192 
\pi^3 f_\pi^4 \sqrt{t}} \int_0^{Q/2} dk\, {\sqrt{m_\pi^2+k^2} \over k} \ln
{k+\sqrt{m_\pi^2+k^2} \over m_\pi}  \nonumber \\ && +{1\over 9\pi^3 (4f_\pi)^4}
\bigg\{ {g_A^4 t \over 20}(20m_\pi^2-7t)+{g_A^2 Q^4\over 4t}(8m_\pi^2-5t)+{16
m_\pi^6 \over t} -12m_\pi^4 +{t^2 \over 5} \bigg\} \nonumber \\ && \times \ln{ 
\sqrt{t}+Q\over 2m_\pi} +{Q \over (24\pi)^3 f_\pi^4 \sqrt{t} } \bigg\{{g_A^4 
\over 100} (257t^2+2604 m_\pi^2 t-6368 m_\pi^4)\nonumber \\ &&+ {g_A^2 \over 4}
Q^2(13t-48m_\pi^2) +{1\over 25}(684 m_\pi^4+73m_\pi^2t -16t^2) \bigg\}
\nonumber \\ &&+{Q^3 \over 480\pi f_\pi^2 \sqrt{t} } \Big[10(\bar d_1+\bar d_2)
(2m_\pi^2-t) -3 \bar d_3 Q^2 +40 \bar d_5 m_\pi^2 \Big] \,.\end{eqnarray}
The term in the last line originates from the third order scale-independent 
low-energy constants $\bar d_j$ introduced in ref.\cite{nadja} which subsume
already chiral logarithms $\ln(m_\pi/\lambda)$. The remainder in eq.(12) is 
identical to the $2\pi$-phase space integral $(8\pi\sqrt{t})^{-1} \int_{-Q/2
}^{Q/2} dk\, {\rm Re}\,[-ik \,g^-(ik,t)]$ with $g^-(\omega,t)$ the isospin-odd
non-spin-flip one-loop $\pi N$-amplitude \cite{nadja,pin}. The analogous 
contribution to the isovector magnetic spectral function reads:   
\begin{eqnarray} {\rm Im}\, G_M^V(t) &=& {g_A^2 M\over \pi^2 (4f_\pi)^4 
\sqrt{t}} \bigg\{ g_A^2m_\pi^2(5m_\pi^2-t)\ln{\sqrt{t} +Q\over 2 m_\pi}  - Q^3
{m_\pi \over 6}\nonumber \\ && + {Q^5 \over 24 \sqrt{t}}\ln{\sqrt{t} +2m_\pi
\over \sqrt{t}-2m_\pi} +{g_A^2 Q \over 24} (30m_\pi^2\sqrt{t}-64 m_\pi^3-
t^{3/2} ) \bigg\}\,.\end{eqnarray}
Again, this expression is identical to $M(32\pi\sqrt{t})^{-1}\int_{-Q/2}^{Q/2} 
dk\,(Q^2-4k^2) {\rm Re}\, h^-(ik,t)$ with $h^-(\omega,t)$ the isospin-odd 
spin-flip one-loop $\pi N$-amplitude \cite{nadja,pin}. We emphasize the correct
threshold behavior of the two-loop contributions: ${\rm Im}\,\sigma_N(t) \sim 
Q$ according to eq.(11) and ${\rm Im}\,G_{E,M}^V(t) \sim Q^3$ according to 
eqs.(12,13). In the latter two cases one easily verifies that the coefficients
of $Q$ vanish identically.   
 
The full set of two-loop graphs contains the two factorizable diagrams
exhibited in Fig.\,2. Both cuts of the intermediate pion-pair generate a
contribution to the imaginary parts, according to the rule: ${\rm
Im}(z_1z_2)={\rm Re}\,z_1 \cdot {\rm Im} \,z_2 + {\rm Im}\,z_1 \cdot{\rm
Re}\,z_2$. Only one term of this sum is accounted for in the expressions
eqs.(11,12,13). The other term arises from that cut which splits off the
one-loop scalar or charge form factor of the pion \cite{gasleut} (as
symbolized by the vertical dotted line in Fig.\,2). Naturally, the scalar form 
factor of the pion enters the scalar spectral function:  
\begin{equation}  {\rm Im}\,\sigma_N(t)  =  {g_A^2 m_\pi^2(t-2 m_\pi^2) \over
\pi^3 (4f_\pi)^4 \sqrt{t}} A(t) \bigg\{2\pi^2 f_\pi^2\langle r^2_S\rangle_\pi 
\,t+ {25t\over 16} -{3\over 8}m_\pi^2(1+\bar l_3) + {3Q\over 4\sqrt{t}}
(m_\pi^2-2t) \ln{ \sqrt{t}+Q\over 2m_\pi}\bigg\}\,.\end{equation}
Here, $\langle r^2_S \rangle_\pi$ denotes the mean square scalar radius of the
pion. The low-energy constant $\bar l_3 \simeq 3$ \cite{gasleut} shows up in 
the one-loop representation of the pionic sigma-term: $\sigma_\pi(0) = \hat m\,
\partial m_\pi^2 /\partial \hat m = m_\pi^2+m_\pi^4(1-\bar l_3)/32\pi^2 
f_\pi^2 \simeq 0.99 m_\pi^2$ \cite{gasleut}. On the other hand side, 
the (one-loop) pion charge form factor \cite{gasleut} enters the
isovector electric and magnetic spectral functions:   
\begin{eqnarray} {\rm Im}\, G_E^V(t) &=& {1\over 9\pi^3 (4f_\pi)^4 \sqrt{t}}
\bigg\{ Q\Big[ g_A^2(5t-8m_\pi^2)+Q^2 \Big] -{3g_A^2 \over M}(t-2m_\pi^2)^2 
A(t) \bigg\} \nonumber \\ && \times \bigg\{2 \pi^2 f_\pi^2 \langle r^2_{ch} 
\rangle_\pi \,t+{t\over 3}-m_\pi^2  -{Q^3 \over 4\sqrt{t}} \ln{ \sqrt{t}+
Q\over 2m_\pi}\bigg\}\,,\end{eqnarray} 
\begin{eqnarray} {\rm Im}\, G_M^V(t) &=& {g_A^2 M Q^2 A(t)\over 3\pi^3 (4f_\pi
)^4 \sqrt{t}}\bigg\{(2\pi f_\pi)^2 \langle r^2_{ch}  \rangle_\pi\, t +{2t\over
3} -2m_\pi^2  -{Q^3 \over 2\sqrt{t}} \ln{ \sqrt{t}+Q\over 2m_\pi}\bigg\}
\,,\end{eqnarray}
where  $\langle r^2_{ch}  \rangle_\pi$ denotes the mean square charge radius of
the pion. In order to guarantee the correct threshold behavior Im\,$G_E^V(t)
\sim Q^3$ we have kept the first relativistic $1/M$-correction (see eq.(7)) in 
the first factor of eq.(15). 

\bigskip

\bild{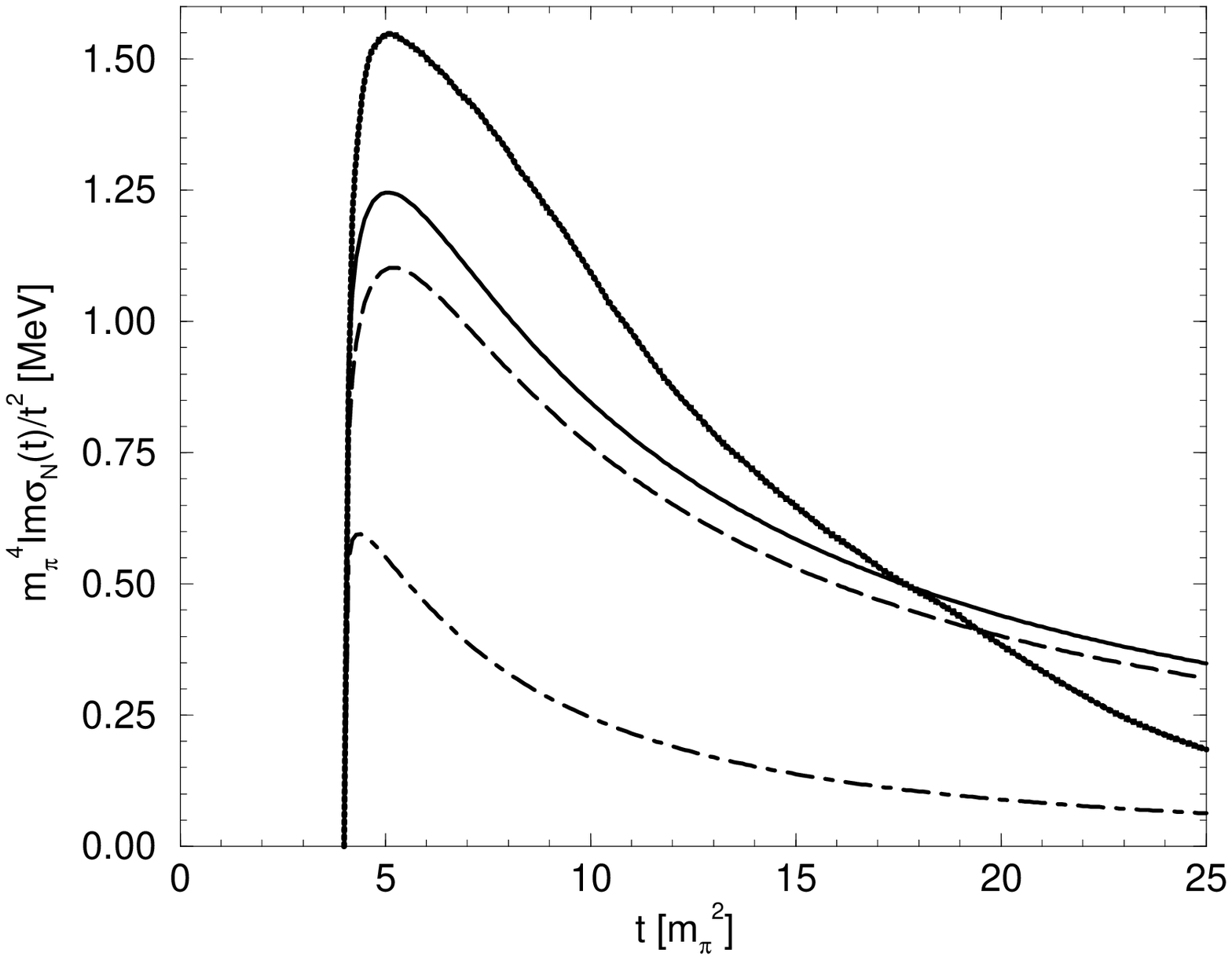}{14}
\vspace{-1.0cm}
{\it Fig.3: The spectral function Im\,$\sigma_N(t)$ of the isoscalar scalar 
form factor of the nucleon multiplied with $m_\pi^4/t^2$. The dashed-dotted
line gives the one-loop result eq.(5). The dashed line includes in addition the
$c_{1,2,3}$-term eq.(9) and the full line includes furthermore the two-loop
contributions eqs.(11,14). The upper dotted line shows the empirical spectral
function  $m_\pi^4$\,Im\,$\sigma_N(t)/t^2$ of ref.\cite{sigmaterm}.}

\bigskip
Let us add a remark on chiral power counting of the spectral functions. 
Counting the quantities $\sqrt t$, $m_\pi$ and $Q=\sqrt{t-4m_\pi^2}$ as small
momenta one deduces that two-loop contributions to Im\,$\sigma_N(t)$, 
Im\,$G_E^V(t)$ and  Im\,$G_M^V(t)$ are of fifth, fourth and third order in 
small momenta, respectively. In the latter two cases the electric/magnetic 
coupling of the (virtual) photon absorbs one/two powers of small momenta.  

Let us now turn to numerical results. We use the parameters: $m_\pi =
139.57\,$MeV (charged pion mass), $f_\pi = 92.4\,$MeV (pion decay constant) and
$g_A=1.3$ (equivalent to a $\pi N$-coupling constant $g_{\pi N} = g_A M/f_\pi 
= 13.2$ \cite{pavan}). For the second and third order low-energy constants 
$c_i$ and $\bar d_j$ we choose the average values of three fits to $\pi N
$-phase shift solutions given in ref.\cite{nadja}: $c_1=-1.4\,{\rm GeV}^{-1}$, 
$c_2 =3.2\, {\rm GeV}^{-1}$, $c_3=-6.0\,{\rm GeV}^{-1}$, $c_4= 3.5\,{\rm 
GeV}^{-1}$, $\bar d_1+\bar d_2= 3.0\,{\rm GeV}^{-2}$, $\bar d_3 = -3.0\,{\rm
GeV}^{-2}$ and $\bar d_5= 0.1\,{\rm GeV}^{-2}$. Most of these values are
compatible with the low-energy constants $c_i$ and $\bar d_j$ obtained in 
ref.\cite{buett} in a one-loop analysis of $\pi N$-scattering inside the
Mandelstam triangle. The mean square charge radius of the pion is accurately
measured in elastic pion-electron scattering: $\langle r^2_{ch} \rangle_\pi
=(0.44\pm 0.01)\,{\rm fm}^2$ \cite{amendolia}. A value compatible with that has
been found in ref.\cite{bijn} in a two-loop analysis of the low-energy 
$\pi^-e^-$-scattering data. An improved determination of the mean square scalar
radius of the pion has recently been given in ref.\cite{colangelo}: $\langle 
r^2_S \rangle_\pi = (0.61 \pm 0.04)\,{\rm fm}^2$. We use throughout the central
values of these empirical pion mean square radii.    

\bigskip

\bild{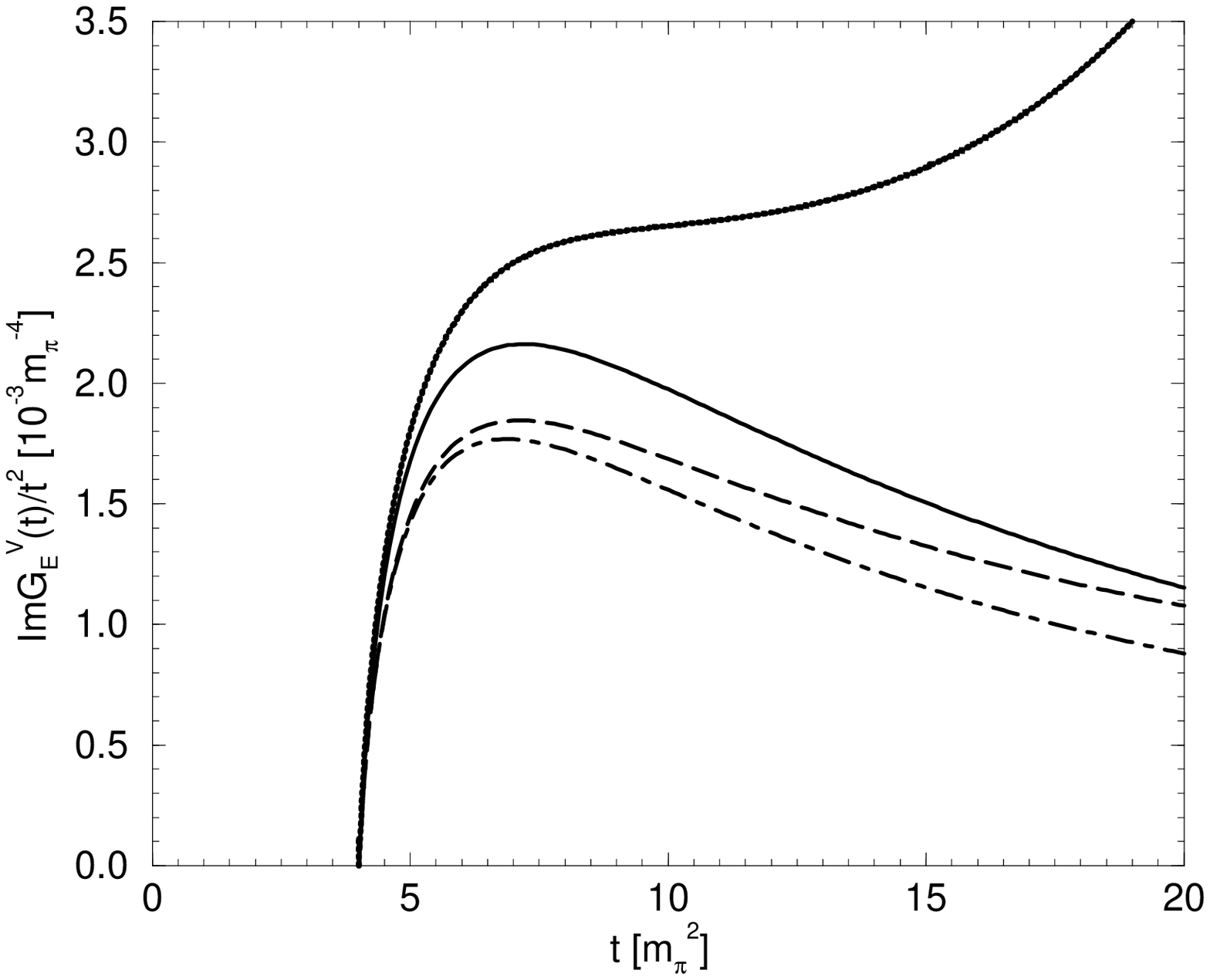}{14}
\vspace{-1.0cm}
{\it Fig.4: The spectral function Im\,$G_E^V(t)$ of the isovector electric 
form factor of the nucleon divided by $t^2$. The dashed-dotted line
gives the one-loop result eq.(7). The dashed line includes in addition the
$c_4/M$-term eq.(10) and the full line includes furthermore the two-loop
contributions eqs.(12,15). The upper dotted line shows the empirical spectral
function  Im\,$G_E^V(t)/t^2$ of ref.\cite{hoehler}.}

\bigskip

In Fig.\,3 we show the spectral function Im\,$\sigma_N(t)$ of the isoscalar
scalar form factor of the nucleon multiplied with a weighting factor 
$m_\pi^4/t^2$. The dashed-dotted line gives the one-loop result eq.(5). The 
dashed line includes in addition the $c_{1,2,3}$-term eq.(9) and the full line
includes furthermore the two-loop contributions eqs.(11,14). The upper dotted
line shows the empirical isoscalar scalar spectral function $m_\pi^4$\,Im\,$
\sigma_N(t)/t^2$ of ref.\cite{sigmaterm}. It has been determined on the basis 
of the $\pi\pi \to \bar NN$ s-wave amplitude $f_+^0(t)$ \cite{hoehler} and 
$\pi\pi$-scattering data tied together with Roy-equations etc. (see eq.(3) in
ref.\cite{sigmaterm}). One observes a substantial improvement when including
the next-to-leading order $c_{1,2,3}$-term and the two-loop contributions. The
major effect comes evidently from the large isoscalar $\pi\pi NN$-contact
couplings, in particular from $c_3$. The height of the peak at $t \simeq
5m_\pi^2$ is however still underestimated by about $20\%$ in two-loop
approximation. Higher order $\pi\pi$-rescattering effects etc. are necessary in
order to close this remaining gap. Given the pattern in Fig.\,3 one can expect
significant effects (in the right direction) already from the two-loop diagrams
with one vertex proportional to the (numerically large) second order low-energy
constants $c_{1,2,3,4}$. Note that a complete calculation of elastic $\pi N
$-scattering to chiral order four which includes the pertinent one-loop 
diagrams with one $c_{1,2,3,4}$-vertex has been recently performed in
ref.\cite{order4}. For comparison, similar deficiencies of the two-loop
approximation of chiral perturbation theory have been observed in 
ref.\cite{pionffs} for the imaginary parts of the pion scalar and charge form 
factors.

Next, we show in Fig.\,4 the spectral function Im\,$G_E^V(t)$ of the isovector 
electric form factor of the nucleon weighted with $1/t^2$. The dashed-dotted 
line gives the one-loop result eq.(7). The dashed line includes in addition the
$c_4/M$-term in eq.(10) and the full line includes furthermore the two-loop
contributions eqs.(12,15). The upper dotted line corresponds to the empirical 
isovector electric spectral function  Im\,$G_E^V(t)/t^2$ of ref.\cite{hoehler}.
Modulo a kinematical factor $Q^3/8M\sqrt{t}$ it is determined by the product of
the $\pi\pi\to \bar NN$ p-wave amplitude $f_+^1(t)$ \cite{hoehler} and the
(time-like) pion charge form factor measured in the reaction $e^+e^- \to \pi^+
\pi^-$ (see eq.(7) in ref.\cite{spectral}). In the case of the isovector
electric spectral function the two-loop contribution is considerably larger
than the $c_4/M$-term (of the same chiral order). The peak at low
$\pi\pi$-invariant masses, $t\simeq 7m_\pi^2$, gets effectively enhanced by a
factor of about $1.2$ by this additional $\pi N$- and $\pi\pi$-rescattering
dynamics. The other prominent feature of the empirical isovector electric
spectral function Im\,$G_E^V(t)$, namely its rise to the $\rho(770)$-resonance,
is clearly absent in (chiral) perturbation theory.

\bigskip

\bild{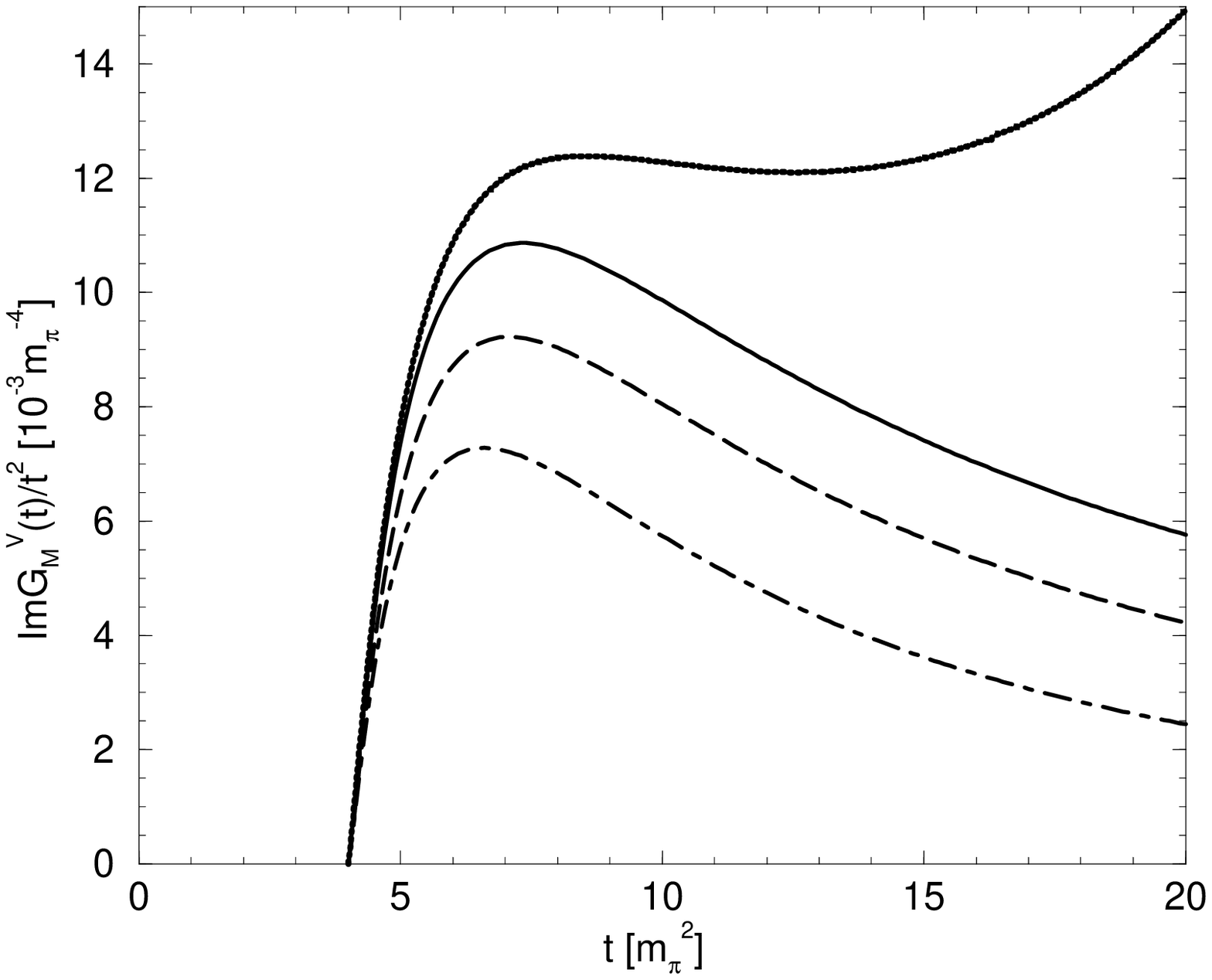}{14}
\vspace{-1.0cm}
{\it Fig.5: The spectral function Im\,$G_M^V(t)$ of the isovector magnetic 
form factor of the nucleon divided by $t^2$. The dashed-dotted line
gives the one-loop result eq.(8). The dashed line includes in addition the
$c_4$-term eq.(10) and the full line includes furthermore the two-loop
contributions eqs.(13,16). The upper dotted line shows the empirical spectral
function  Im\,$G_E^V(t)/t^2$ of ref.\cite{hoehler}.} 

\bigskip

Furthermore, we show in Fig.\,5 the spectral function Im\,$G_M^V(t)$ of the 
isovector magnetic form factor of the nucleon weighted with $1/t^2$. The 
dashed-dotted line gives the one-loop result eq.(8). The dashed line includes 
in addition the $c_4$-term in eq.(10) and the full line includes furthermore
the two-loop contributions eqs.(13,16). The upper dotted line corresponds to
the empirical isovector magnetic spectral function  Im\,$G_M^V(t)/t^2$ of 
ref.\cite{hoehler}. Modulo a kinematical factor $Q^3/8\sqrt{2t}$ it is 
determined by the product of the $\pi\pi\to\bar NN$ p-wave amplitude $f_-^1(t)$
and the (time-like) pion charge form factor (see again eq.(7) in
ref.\cite{spectral}). In the case of the isovector magnetic spectral function
the two-loop contribution and the next-to-leading order $c_4$-term are
approximately equal in magnitude. Both together enhance the peak slightly above
threshold considerably by a factor of about $1.5$. Again, the strong rise of
the empirical isovector magnetic spectral function Im\,$G_M^V(t)$ to the
$\rho(770)$-resonance cannot be reproduced in (chiral) perturbation theory. 

For a more complete description of the dynamics governing the empirical 
isovector electromagnetic spectral functions the introduction of an explicit 
$\rho(770)$-resonance is indispensable. For that reason we add a 
phenomenological $\rho$-meson exchange contribution of the simple form: 
\begin{equation} {\rm Im}\,G_{E,M}^V(t) = {b_{E,M}\, m_\rho^2 \sqrt{t}\,
\Gamma_\rho(t) \over (m_\rho^2-t)^2+t \,\Gamma_\rho^2(t)}\,, \qquad
\Gamma_\rho(t) = {g_{\rho \pi}^2 Q^3 \over 48 \pi t} \,, \end{equation}
which of course includes the proper energy-dependent $\rho\to 2\pi$ decay width
$\Gamma_\rho(t)$. We take the value $m_\rho = 769.3\,$MeV \cite{pdg} for the 
$\rho$-meson mass and coupling constant $g_{\rho \pi}=6.03$ is determined from 
the empirical decay width $\Gamma_\rho(m_\rho^2)=150.2\,$MeV \cite{pdg}. 
Furthermore, $b_E = 1.0$ and $b_M = 3.5$ are numerical parameters which we have
adjusted to the height of the $\rho(770)$-resonance peak (at $t\simeq 28
m_\pi^2$) of the empirical isovector electric and magnetic spectral functions 
Im\,$G_{E,M}^V(t)/t^2$.

\bigskip

\bild{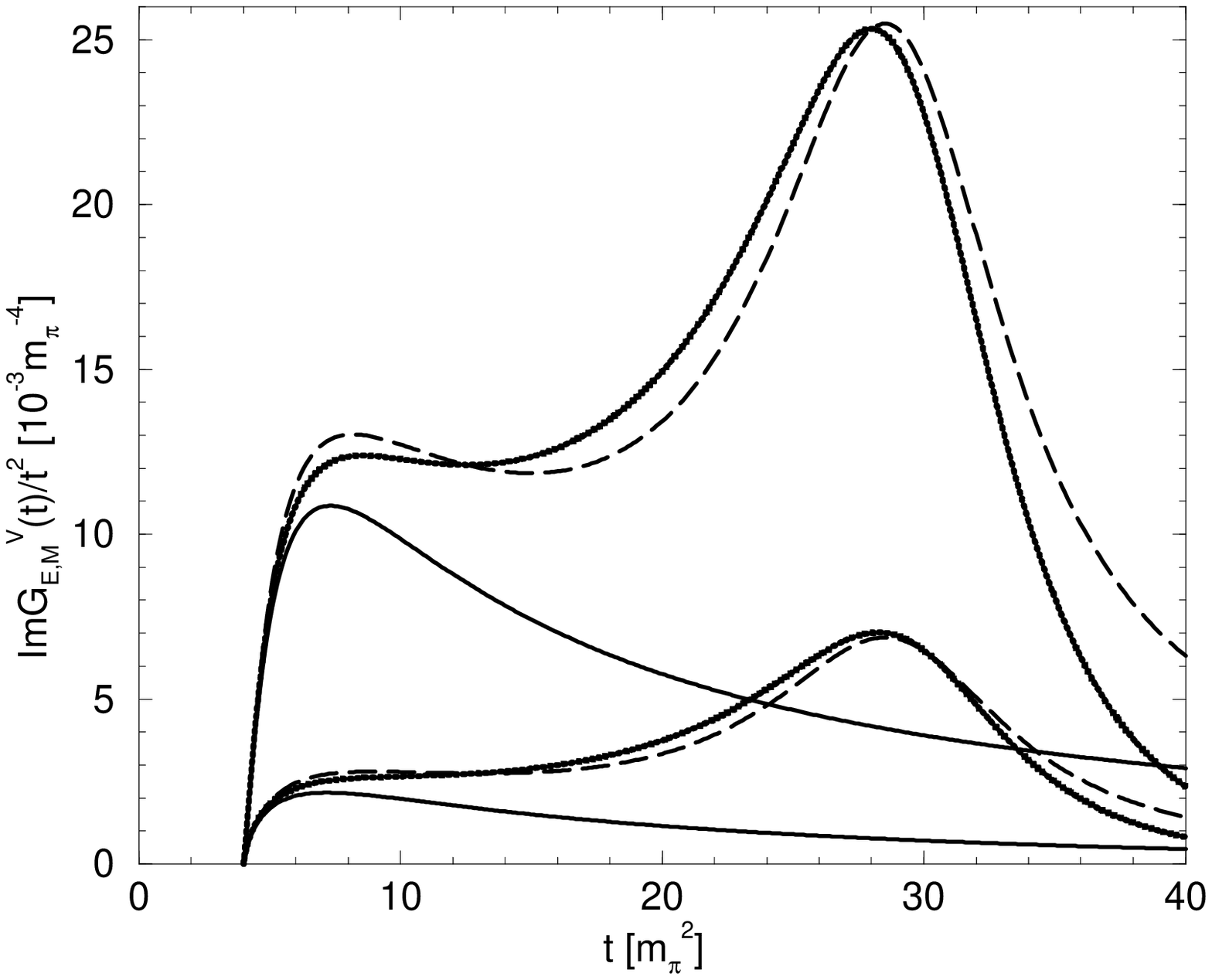}{14}
\vspace{-1.0cm}
{\it Fig.6: Spectral functions of the isovector electric and magnetic form 
factor of the nucleon weighted with $1/t^2$. Shown are Im\,$G_E^V(t)/t^2$ 
(lower three lines) and Im\,$G_M^V(t)/t^2$ (upper three lines). The full lines
give the two-loop results of chiral perturbation theory and the dashed lines 
include in addition the phenomenological $\rho$-meson contribution eq.(17). The
dotted curves correspond to the empirical spectral function 
Im\,$G_{E,M}^V(t)/t^2$ of ref.\cite{hoehler}.} 

\bigskip

Finally, the lower (upper) lines in Fig.\,6 show the isovector electric 
(magnetic) spectral function weighted with $1/t^2$. The full lines give the 
respective two-loop result of chiral perturbation theory and the dashed lines 
include in addition the phenomenological $\rho$-meson exchange contribution 
eq.(17). One observes that the empirical isovector electric and magnetic
spectral functions (dotted lines) can now be reproduced fairly well. The
agreement is rather satisfactory for the electric one. In the case of the
isovector magnetic spectral function Im\,$G_M^V(t)/t^2$ some overshooting and
undershooting of the empirical values below and above $t \simeq 13m_\pi^2$ is 
visible in Fig.\,6. This may point towards the relevance of higher order
$\pi\pi$-rescattering effects, or it may just result from the simplicity of the
$\rho(770)$-resonance parameterization eq.(17). In any case one should not
overinterpret the small differences between the dashed and dotted lines in
Fig.\,6 since there may be a potential danger of double-counting when adding 
together perturbative chiral pion loops and a finite width $\rho(770)$-meson 
exchange term (representing non-perturbative pion-dynamics). 

In summary, we have calculated in this work the imaginary parts of the 
isoscalar scalar and isovector electromagnetic form factors of the nucleon up 
to two-loop order in chiral perturbation theory. We have performed the
non-relativistic $1/M$-expansion such that the correct threshold behavior, 
Im\,$\sigma_N(t)\sim \sqrt{t-4m_\pi^2} $ and Im\,$G_{E,M}^V(t)\sim(t-4m_\pi^2
)^{3/2}$, is ensured. The two-loop contributions and the next-to-leading order
$c_{1,2,3,4}$-terms magnify the strong enhancement slightly above the two-pion
threshold. Higher order $\pi\pi$-rescattering effects etc. are still necessary
to close a remaining $20\%$-gap to the empirical isoscalar scalar spectral 
function Im\,$\sigma_N(t)$ \cite{sigmaterm}. After adding a phenomenological
$\rho(770)$-meson exchange term to the respective two-loop results the 
empirical isovector electric and magnetic spectral functions Im\,$G_{E,M}^V(t)$
\cite{hoehler} can be reproduced fairly well. 

\bigskip

I thank Ulf-G. Mei{\ss}ner for his tabulated values of the empirical isovector
electromagnetic spectral functions Im\,$G_{E,M}^V(t)$ and for critical reading
of the manuscript.


\begin{thebibliography}{99} 
\bibitem{fulco} W.R Frazer and J.R. Fulco, {\it Phys. Rev. Lett.} {\bf 2}, 365 
(1959); {\it Phys. Rev.} {\bf 117}, 1609 (1960).\vs
\bibitem{hoehler} G. H\"ohler and E. Pietarinen, {\it Phys. Lett.} {\bf B53}, 
471 (1975); G. H\"ohler, Landolt-B\"ornstein, Vol.9b2, ed. H. Schopper,
Springer-Verlag (1983), chapters 2.5.2 and 11.\vs
\bibitem{mergel} P. Mergell, Ulf-G. Mei{\ss}ner and D. Drechsel, {\it Nucl. 
Phys.} {\bf A596}, 367 (1996); and references therein.\vs
\bibitem{hammer} H.W. Hammer, Ulf-G. Mei{\ss}ner and D. Drechsel, {\it Phys. 
Lett.} {\bf B385}, 343 (1996).\vs
\bibitem{gss} J. Gasser, M.E. Sainio and A. Svarc, {\it Nucl. Phys.} {\bf
B307}, 779 (1988).\vs
\bibitem{spectral} V. Bernard, N. Kaiser and Ulf-G. Mei{\ss}ner, {\it Nucl. 
Phys.} {\bf A611}, 429 (1996).\vs
\bibitem{sigmaterm} J. Gasser, H. Leutwyler and M.E. Sainio, {\it Phys. Lett.}
{\bf B253}, 260 (1991); and references therein.\vs 
\bibitem{infrared} B. Kubis and Ulf-G. Mei{\ss}ner, {\it Nucl. Phys.} {\bf 
A679}, 698 (2001).\vs
\bibitem{nadja} N. Fettes, Ulf-G. Mei{\ss}ner and S. Steininger, {\it Nucl. 
Phys.} {\bf A640}, 199 (1998).\vs
\bibitem{pin} V. Bernard, N. Kaiser and Ulf-G. Mei{\ss}ner, {\it Nucl. 
Phys.} {\bf A615}, 483 (1997).\vs
\bibitem{gasleut} J. Gasser and H. Leutwyler, {\it Ann. Phys.} {\bf 158}, 142 
(1984).\vs 
\bibitem{buett} P. B\"uttiker and Ulf-G. Mei{\ss}ner, {\it Nucl. Phys.} {\bf 
A668}, 97 (2000).\vs
\bibitem{pavan} M.M. Pavan, R.A. Arndt, I.I. Strakovsky and R.L. Workman, {\it
Phys. Scr.} {\bf T87}, 65 (2000).\vs 
\bibitem{amendolia} S.R. Amendolia et al., NA7 Collaboration, {\it Nucl. Phys.}
{\bf B277}, 168 (1986).\vs
\bibitem{bijn} J. Bijnens and P. Talavera, {\it JHEP} {\bf 0203} (2002) 046.\vs
\bibitem{colangelo} G. Colangelo, J. Gasser and H. Leutwyler, {\it Nucl. Phys.}
{\bf B603}, 125 (2001); and references therein.\vs
\bibitem{order4} N. Fettes and Ulf-G. Mei{\ss}ner, {\it Nucl. Phys.} {\bf
A676}, 311 (2000).\vs 
\bibitem{pionffs} J. Gasser and Ulf-G. Mei{\ss}ner, {\it Nucl. Phys.} {\bf
B357}, 90 (1991).\vs 
\bibitem{pdg} Review of Particle Properties, D.E. Groom et al., {\it
Eur. Phys. J.} {\bf C15}, 1 (2000).\vs
\end{thebibliography}
\end{document}